\documentclass[conference]{IEEEtran}
\IEEEoverridecommandlockouts
\usepackage{hyperref}
\usepackage[style=authoryear,natbib=true,backend=biber,backref=true,maxcitenames=3,maxbibnames=99,babel=other]{biblatex}
\usepackage{amsmath,amssymb,amsfonts}
\usepackage{algorithmic}
\usepackage{graphicx}
\usepackage{textcomp}
\usepackage{mathtools}
\usepackage{booktabs}
\usepackage{xcolor}
\definecolor{aquali}{cmyk}{1,0.0474,0,0.255} 
\hypersetup{colorlinks=true,urlcolor=aquali,linkcolor=black,citecolor=blue}
\providecommand{\tightlist}{%
  \setlength{\itemsep}{0pt}\setlength{\parskip}{0pt}}

\def\EtienneSimon{\'Etienne Simon}
\def\KevinCocchi{K\'evin Cocchi}
\def\Sedeto{Sol\`ene Pichereau}

\begin{document}

\title{Using Posters to Recommend Anime and Mangas in a Cold-Start Scenario}

\author{\IEEEauthorblockN{Jill-J\^enn Vie}
\IEEEauthorblockA{\textit{Mangaki} \\
\textit{RIKEN Center for AIP}\\
Tokyo, Japan \\
jill-jenn.vie@riken.jp}
\and
\IEEEauthorblockN{Florian Yger}
\IEEEauthorblockA{\textit{LAMSADE} \\
\textit{Universit\'e Paris-Dauphine}\\
Paris, France\\
florian.yger@dauphine.fr}
\and
\IEEEauthorblockN{Ryan Lahfa, Basile Clement,\\K\'evin Cocchi, Thomas Chalumeau}
\IEEEauthorblockA{\textit{Mangaki}\\
Mitry-Mory, France\\
\{ryan, basile, kevin, thomas\}@mangaki.fr}
\and
\IEEEauthorblockN{Hisashi Kashima}
\IEEEauthorblockA{\textit{Kyoto University} \\
\textit{RIKEN Center for AIP}\\
Kyoto, Japan\\
kashima@i.kyoto-u.ac.jp}
}

\maketitle

\DeclarePairedDelimiter{\norm}{\lVert}{\rVert}

\begin{abstract}
Item cold-start is a classical issue in recommender systems that affects anime and manga recommendations as well. This problem can be framed as follows: how to predict whether a user will like a manga that received few ratings from the community? Content-based techniques can alleviate this issue but require extra information, that is usually expensive to gather. In this paper, we use a deep learning technique, Illustration2Vec, to easily extract tag information from the manga and anime posters (e.g., sword, or ponytail). We propose BALSE (Blended Alternate Least Squares with Explanation), a new model for collaborative filtering, that benefits from this extra information to recommend mangas. We show, using real data from an online manga recommender system called Mangaki, that our model improves substantially the quality of recommendations, especially for less-known manga, and is able to provide an interpretation of the taste of the users.
\end{abstract}


\section{Introduction}\label{introduction}

Recommender systems are useful to help users decide what to enjoy next.
In the case of anime and mangas, users, easily overwhelmed by the
ever-growing amount of works, end up barely scratching the surface of
what Japanese animation has to offer. Collaborative filtering is a
popular technique that relies on existing rating data from users on
items in order to predict unseen ratings
\autocite{aggarwal2016recommender}. However, it is still hard to
recommend items for which little information is available, e.g., items
for which few or no ratings have been provided by the community. This
problem has been referred to as the \emph{item cold-start problem}.

In order to alleviate this problem, it is possible to rely on extra
information about the items, such as metadata (e.g., for movies:
directors, composers, release date). However, such information is not
always available: new anime projects may only have a poster or a
trailer, and a title. Such a poster is usually the first contact that a
user has with an item and plays a large role in the user's decision to
watch it or not. Especially in the manga and anime industry, posters
contain a lot of information about the characters, in order to maximize
the visual appeal for the consumers. Hence, it is natural to consider
posters as a source for additional metadata. In recent years,
convolutional neural networks (CNNs) have established themselves as the
\emph{de-facto} method for extracting semantic information from image
content in a wide variety of tasks. We propose using a CNN for
extracting meaningful tags directly from the item's poster. Such
extracted tags can help draw links between items, which can be useful
when few ratings are available. In this paper, we present
BALSE\footnote{http://knowyourmeme.com/memes/events/balse} (Blended
Alternate Least Squares with Explanation), a new method leveraging tag
information extracted from the posters for tackling the item cold-start
problem and improving the recommendation performance for little-known
items. We show using real data that our method provides better rating
predictions than existing techniques, and gives interpretable insight
about the user's taste.

To the best of our knowledge, this is the first research work that uses
tag prediction on item posters in order to improve the accuracy of a
recommender system and explain to users why they are recommended
little-known items.

This paper is organized as follows. We first present existing work
related to this research. Then, we expose the context of collaborative
filtering and item cold-start, together with a few common assumptions.
We then describe our model, BALSE, and present some experimental results
on a real dataset. We finish by discussing the results and future work.

\section{Related Work}\label{related-work}

Using side information in order to improve recommendations has been the
core of existing research \autocite{kula2015metadata} and several
approaches have been developed to take into account extra data about
users or items, whether coming from text
\autocites{alexandridis2017parvecmf}{fang2011matrix}, social networks
\autocite{delporte2013socially}, images or other types of data
\autocites{nedelec2017specializing}{kim2014scalable}{xu2013speedup}.
More recently, deep learning techniques have been used for this purpose.
YouTube is extracting features from the videos browsed within a user
history in order to improve their recommendations
\autocite{Covington2016}. Researchers have also analyzed music content
as extra information \autocite{Van2013}. They managed to recover
explainable latent features, corresponding to certain types of music,
without any human annotation. Such side information is particularly
useful in order to mitigate the cold-start problem
\autocites{wei2017collaborative}{kula2015metadata}{biswas2017combating}{bobadilla2012collaborative}.
In the exact context of movies, \textcite{Zhao2016} extract latent
features from the posters using CNNs and improve the recommendations
using those latent features. However, those extracted features do not
have semantic meaning, therefore they cannot be used to explain to the
user why extra works are recommended to them.

Several approaches have tried to bridge the gap between content-based
approaches and collaborative filtering \autocite{burke2002hybrid}. The
main idea behind those so-called hybrid methods is to combine different
recommendation models in order to overcome their limitations and build
more robust models with higher predictive power. Existing techniques can
take on several names: blending or stacking
\autocites{roda2011optimal}{jahrer2010combining}, or the general
ensemble methods for machine learning estimators. These techniques use
the output of different models as features for a higher-level model.
This higher-level model is usually a linear model
\autocite{sill2009feature}. Such blended methods have played an
important role in achieving top performance in challenges such as the
Netflix Prize \autocites{sill2009feature}{koren2009bellkor}. The
approach described in this paper builds upon these ideas, as we are
presenting a blended model, but the combination we present is nonlinear.
We complement a classical collaborative filtering recommender system
with a fallback model that will compensate the prediction error on
hard-to-predict data points, i.e.~items with few ratings.

\section{Context}\label{context}

We assume the typical setting for collaborative filtering: we have
access to a \(n \times m\) rating matrix \(R\) containing the ratings of
\(n\) users on \(m\) items that can be either manga or anime: \(r_{ij}\)
represents the rating that user \(i\) gave to item \(j\). In practice,
since users only rate the few items that they have interacted with, the
rating matrix \(R\) tend to be very sparse: in the dataset considered in
this paper, less than 1\% of the entries are known; other popular
datasets in the field
\autocites{aggarwal2016recommender}{wang2016learning} report similar
levels of sparsity. Therefore, it is challenging to infer the missing
entries of \(R\).

Another assumption is that the whole rating matrix can be explained by
few latent profiles, i.e.~each user rating vector can be decomposed as a
combination of few latent vectors. Therefore, matrix completion is
usually performed using matrix factorization: we try to devise a
factorization \(R \approx UV^T\) where a \(n \times r\) matrix \(U\)
represents the user feature vectors and a \(m \times r\) matrix \(V\)
represents the item feature vectors. Once this model is trained, that
is, when the available entries of \(R\) match their counterparts in
\(UV^T\), computing a missing entry \((i, j)\) of the rating matrix
\(R\) is simply performed by looking at the corresponding \((i, j)\)
entry of \(UV^T\), or, equivalently, computing the dot product
\(U_i^T V_j\) where \(U_i\) in the \(i\)-th row of \(U\) and \(V_j\) is
the \(j\)-th row of \(V\).

Finally, we also assume that we have access to the posters of some of
the items. This is all the content we have.

\section{Our Model: BALSE}\label{our-model-balse}

\begin{figure}
\includegraphics[width=\linewidth]{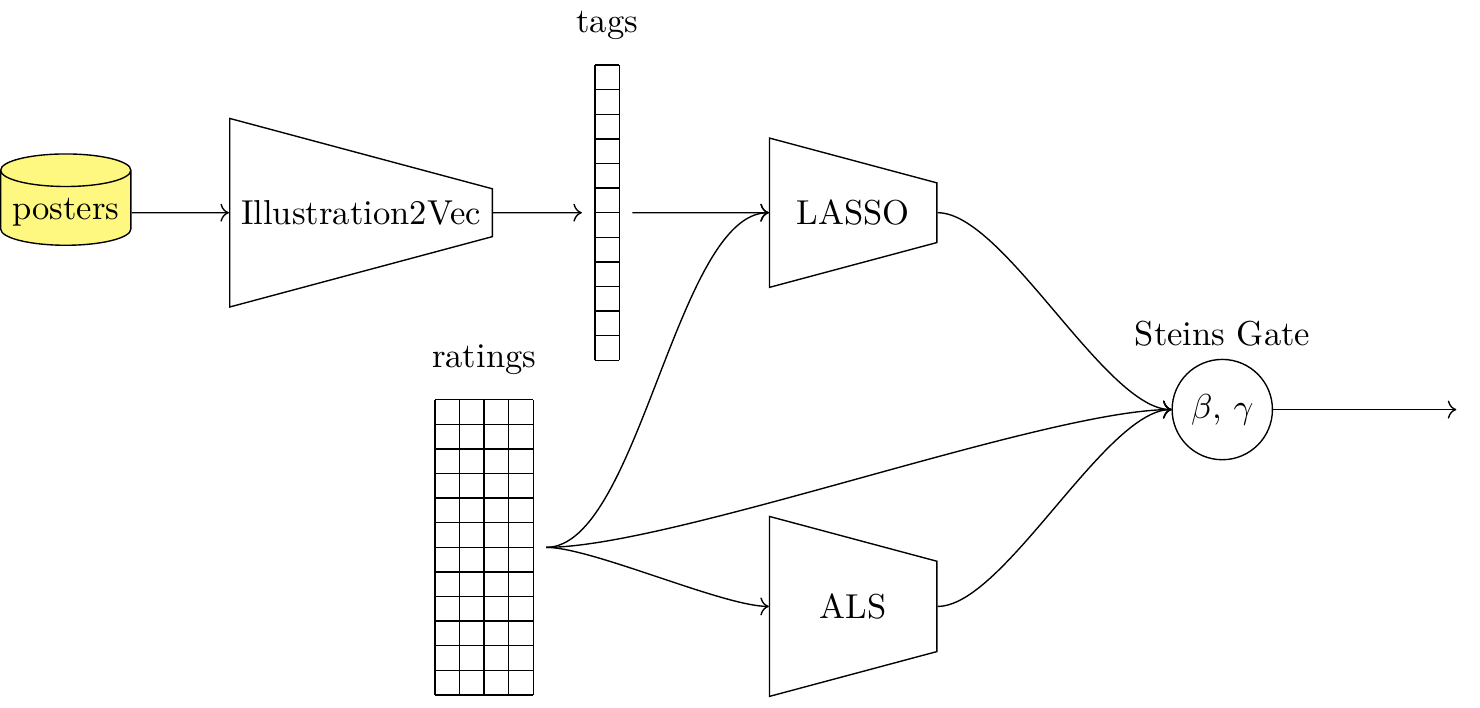}
\caption{The BALSE architecture.}
\label{balse-architecture}
\end{figure}

We now describe BALSE (Blended Alternate Least Squares with
Explanation), our model for recommending anime and mangas. The main idea
is to rely on the rating matrix when possible, and on the posters when
rating information barely exists. We expect a nonlinear blending of two
models considering these sources of information to achieve higher
performance than any of the models. BALSE is composed of several blocks:

\begin{itemize}
\tightlist
\item
  an Illustration2Vec block, which is a convolutional neural network
  that takes a poster as input and outputs tag predictions;
\item
  an ALS\footnote{ALS stands for Alternate Least Squares.} block, that
  performs a matrix factorization of the rating matrix for collaborative
  filtering using alternate least squares with \(\lambda\)-weighted
  regularization;
\item
  a LASSO\footnote{LASSO stands for Least Absolute Shrinkage and
    Selection Operator.} block, that performs a regularized linear
  regression of each row of the rating matrix, using tag predictions, in
  order to infer explainable user preferences;
\item
  a Steins gate, that performs a blending of the outputs of ALS and
  LASSO models, in order to overcome their limitations and provide a
  final rating value.
\end{itemize}

The main architecture of our model is presented in Figure
\ref{balse-architecture}. Both posters and ratings are used for the
predictions.

\subsection{Illustration2Vec}\label{illustration2vec}

\begin{figure*}[h!]
\centering
\includegraphics[height=0.4\linewidth]{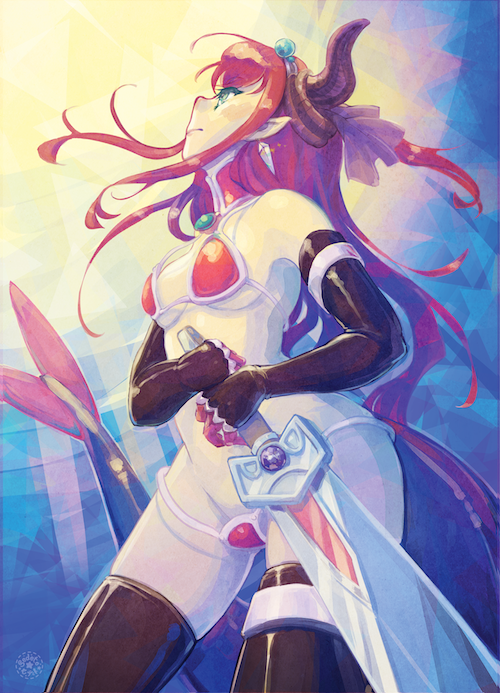}
\includegraphics[height=0.4\linewidth]{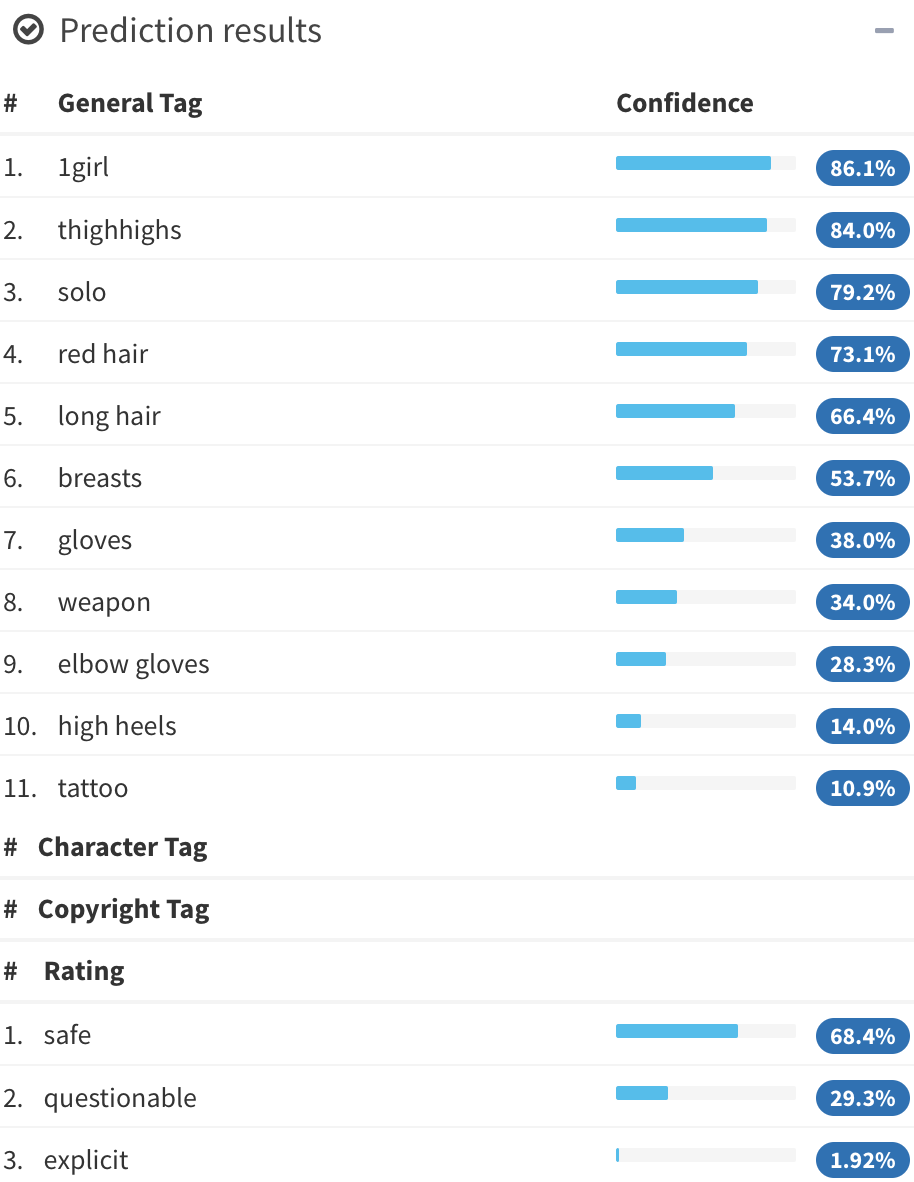}
\caption{Example of tag prediction on a manga illustration. \copyright \Sedeto, http://sedeto.fr.}
\label{illustration2vec}
\end{figure*}

This block extracts tag information from the posters, such as ``1girl''
or ``weapon''. Such tags are associated with confidence weights that
represent how likely a certain tag appears in a certain poster.
Formally, from the poster database, we want to extract a \(m \times t\)
matrix \(T\) where \(m\) is the number of items and \(t\) is the number
of possible tags, such that \(t_{jk} \in [0, 1]\) represents how likely
a tag \(k\) appears in the poster of item \(j\). \(T\) is computed using
Illustration2Vec \autocite{Saito2015}, a VGG-16 neural network
\autocite{simonyan2014very} that predicts a variety of tags based on
illustrations, pre-trained on ImageNet and trained on manga
illustrations labeled with tags from the community website Danbooru. We
use the implementation provided by the authors, which is freely
available. The output of the network is for each poster \(j\), a vector
\(T_j = (t_{j1}, \ldots, t_{jt})\) where for tag \(k = 1, \ldots, t\),
component \(t_{jk} \in [0, 1]\) represents how likely tag \(k\)
describes the poster of item \(j\). In other words, the output of
Illustration2Vec is a row of matrix \(T\). We will call such a vector a
tag prediction. See Fig. \ref{illustration2vec} for an example of an
output of the Illustration2Vec model.

\subsection{LASSO}\label{lasso}

The LASSO block approximates the rating matrix \(R\) with a regularized
linear regression model called LASSO
\autocite{tibshirani1996regression}, using the tag predictions as
features for the items. We train a LASSO model for every user in the
train set.

\begin{equation*}
R \approx P T^T
\end{equation*}

\noindent
where:

\begin{itemize}
\tightlist
\item
  \(P\) contains the parameters to learn, a \(n \times t\) matrix of
  user preferences, of which the \(i\)-th row is denoted as \(P_i\)
  (likewise, \(R_i\) denotes the \(i\)-th row of \(R\));
\item
  \(T\) is the given \(m \times t\) matrix of tag predictions for each
  item.
\end{itemize}

LASSO comes with an \(\alpha\) parameter which induces a L1
regularization term to prevent overfitting, and to provide explanation
of user preferences as we will show later. Therefore, for every user
\(i\) of the train set, we estimate the parameters \(P_i\) that
minimize:

\begin{equation*}
\frac1{2\mathcal{N}_i} \norm{R_i - P_i T^T}_2^2 + \alpha \norm{P_i}_1
\end{equation*}

\noindent
where \(\mathcal{N}_i\) is the number of items rated by user \(i\).

The output of the LASSO block is a rating prediction for each pair
\((i, j)\):

\begin{equation*}
\hat{r}_{ij}^{LASSO} = \tau(P_i^T T_j).
\end{equation*}

\noindent
where \(\tau : x \mapsto \max(\min(x, 2), -2)\) is a function that
shrinks its input to values between -2 and 2. Such a function prevents
the regressor from providing invalid predictions that are outside the
range of rating values.

\subsection{ALS}\label{als}

The ALS block performs matrix factorization of the \(n \times m\) sparse
rating matrix \(R\), in order to provide an estimate
\(\hat{r}_{ij}^{ALS}\) for the missing entries \((i, j)\). Thus, we
learn the parameters of the following factorization:

\begin{equation*}
R \approx U V^T
\end{equation*}

\noindent
where:

\begin{itemize}
\tightlist
\item
  \(U\) is the \(n \times r\) matrix of user latent vectors;
\item
  \(V\) is the \(m \times r\) matrix of item latent vectors.
\end{itemize}

In order to avoid overfitting, we regularize the parameters to estimate.
Therefore, as we want to minimize the squared error, the loss function
to minimize has the following form:

\begin{equation*}
\sum_{i, j | r_{ij} \neq 0} (r_{ij} - U_i^T V_j)^2 + \lambda \left( \norm{U_i}_2^2 + \norm{V_j}^2_2 \right)
\end{equation*}

\noindent where \(U_i\) for every \(i = 1, \ldots, n\) are the rows of
\(U\) and \(V_j\) for every \(j = 1, \ldots, m\) are the rows of \(V\),
and \(\lambda\) is a regularization parameter. This estimation is made
by using alternate least squares with weighted
\(\lambda\)-regularization (ALS-WR) \autocite{Zhou2008}.

Once the parameters have been learned, the prediction for rating of user
\(i\) on item \(j\) is:

\begin{equation*}
\hat{r}_{ij}^{ALS} = U_i^T V_j.
\end{equation*}

\subsection{Steins Gate}\label{steins-gate}

At this step, we have predictions from two different blocks: ALS trained
on the ratings and LASSO trained on the tag predictions of the posters.
We want to improve the predictive power of the overall model, thus we
learn a rule that would automatically choose the best model according to
the number of ratings of the item considered. Formally, we want to learn
parameters \(\beta\) and \(\gamma\) such that:

\begin{align*}
\hat{r}_{ij}^{BALSE} & = \sigma(\beta(\mathcal{R}_j - \gamma)) \hat{r}_{ij}^{ALS}\\
& \quad + \left(1 - \sigma(\beta(\mathcal{R}_j - \gamma))\right) \hat{r}_{ij}^{LASSO}
\end{align*}

\noindent
where:

\begin{itemize}
\tightlist
\item
  \(\mathcal{R}_j\) is the number of ratings of the item \(j\);
\item
  \(\hat{r}_{ij}^{ALS}\) is the rating prediction of ALS model for user
  \(i\) on item \(j\);
\item
  \(\hat{r}_{ij}^{LASSO}\) is the rating prediction of LASSO model for
  user \(i\) on item \(j\);
\item
  \(\sigma : x \mapsto 1/(1 + e^{-x})\) is the sigmoid function.
\end{itemize}

The intuition behind this formula is the following: we want to find a
threshold \(\gamma\) such that when the number of ratings of item \(j\)
verifies \(R_j \gg \gamma\), BALSE mimics ALS, e.g.,
\(\hat{r}_{ij}^{BALSE} \approx \hat{r}_{ij}^{ALS}\), while when
\(R_j \ll \gamma\), i.e.~in a cold-start setting, BALSE mimics LASSO,
e.g., \(\hat{r}_{ij}^{BALSE} \approx \hat{r}_{ij}^{LASSO}\). \(\beta\)
is just a scaling parameter that indicates how sharp the passage from
LASSO to ALS will be. Formally, we want to estimate the parameters
\(\beta\) and \(\gamma\) that minimize:

\begin{equation*}
\sum_{i, j | r_{ij} \neq 0} \left(\hat{r}_{ij}^{BALSE} - r_{ij}\right)^2.
\end{equation*}

This formula is differentiable with respect to \(\gamma\), thus it makes
its optimization easier. It can be seen as a soft switch between the two
possible predictions (ALS and LASSO), according to the number of ratings
of the item. The parameters \(\beta\) and \(\gamma\) are learned using
gradient descent.

\section{Experiments and Results}\label{experiments-and-results}

\subsection{Mangaki dataset}\label{mangaki-dataset}

Mangaki\footnote{https://mangaki.fr} \autocite{Vie2015} is a website
where people can rate items that represent either manga or anime, and
receive recommendations based on their ratings. Mangaki can be seen as
an open source version of Movielens \autocite{movielens2015dataset} for
manga and anime. The Mangaki dataset is a \(2079 \times 9979\)
anonymized matrix of 334390 ratings from 2079 users on 9979 items. 80\%
of the items have a poster.

Users can either rate an item with \{favorite, like, neutral, dislike\}
if they watched it, or \{willsee, wontsee\} if they did not watch it,
i.e.~testify whether they want to watch it or not, based on the content
presented: poster, possibly synopsis, or some statistics.

\subsection{Models}\label{models}

The models considered in this benchmark are:

\begin{itemize}
\tightlist
\item
  ALS: alternate least squares with weighted \(\lambda\)-regularization
  from \autocite{Zhou2008}, that ignores posters;
\item
  LASSO: regularized linear regression using ratings and the tag
  predictions from Illustration2Vec, that is content-based;
\item
  BALSE: the proposed method.
\end{itemize}

In practice, we use \(\lambda = 0.1\) and rank \(r = 20\) for every ALS
model trained and \(\alpha = 0.01\) for every LASSO model trained.
Ratings are mapped into custom values: (favorite, like, neutral,
dislike) = (4, 2, 0.1, -2) and (willsee, wontsee) = (0.5, -0.5). The
Steins gate is optimized using gradient descent with exponential decay
implemented in TensorFlow. The learning rate starts at \(0.9\) and
decays every 20 steps with a base of \(0.997\). All the code is
available on our GitHub repository\footnote{https://github.com/mangaki/balse}.

\subsection{5-fold cross validation}\label{fold-cross-validation}

We perform a 5-fold cross validation over the triplets
\((i, j, r_{ij})\) of the database, keeping 30\% of the train set as a
validation set. Therefore, our data is split into a train set (56\%), a
validation set (24\%) and a test set (20\%).

The vanilla models ALS and LASSO are trained on both the train set and
the validation set. For BALSE, the ALS and LASSO blocks are first
trained using the train set only, and the Steins gate parameters
\(\beta\) and \(\gamma\) are trained using the validation set\footnote{Please
  also note that in Steins gate, the number of ratings \(R_j\) of item
  \(j\) is computed over the train set.}, in order to prevent
overfitting. For the final predictions of BALSE, blending is performed
using the learned \(\beta\) and \(\gamma\) parameters, and the vanilla
ALS and LASSO models. Finally, the root mean squared error (RMSE) is
computed over the test set.

We distinguish the performance of all three models on three sets: the
whole test set, a set of little-known items that received less than 3
ratings in the train and validation set (that represents 1000 ratings,
therefore 3\% of the test set), and cold-start items, i.e.~items that
were never seen in the train and validation sets.

\subsection{Results}\label{results}

BALSE achieves a comparable performance than ALS overall, but
substantially improves the recommendations on little-known items, see
Table \ref{tab-results}.

\begin{table*}
\centering
\caption{Results of RMSE on various subsets of the test set.}
\begin{tabular}{cccc} \toprule
RMSE & Whole test set & Little-known items (3\% least rated) & Cold-start items (0 rating)\\ \midrule
ALS & 1.15681 $\pm$ 0.004 & 1.29269 $\pm$ 0.029 & 1.50047 $\pm$ 0.03500\\
LASSO & 1.44444 $\pm$ 0.002 & 1.31331 $\pm$ 0.036 & 1.37909 $\pm$ 0.05600\\
BALSE & \textbf{1.14954 $\pm$ 0.004} & \textbf{1.22714 $\pm$ 0.036} & \textbf{1.34533 $\pm$ 0.04500}\\ \bottomrule
\end{tabular}
\label{tab-results}
\end{table*}

\begin{figure}
\centering
\includegraphics[width=\linewidth]{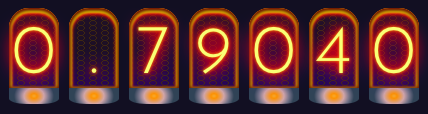}
\caption{This is Steins gate's choice: $\gamma = 0.79040$. For items having at least one rating, it is better to rely more on the ratings predicted by ALS than by LASSO.}
\label{fig-gamma}
\end{figure}

The learned parameter \(\gamma\) of the Steins gate was less than 1, see
Figure \ref{fig-gamma}, which means that items having at least 1 rating
can start to rely on ALS (their ratings) more than LASSO (their poster)
for the predictions. However, BALSE provides better predictions than ALS
for cold-start items, because ALS was not trained on them in the train
set, therefore it outputs constant predictions.

ALS converges after 10 iterations. Steins gate takes 15k iterations to
converge. LASSO is the bottleneck of the proposed approach because one
LASSO model should be trained per user that appears in the train set.

\subsection{Explanation of user taste}\label{explanation-of-user-taste}

Using the tags, it is possible to provide an explanation of the taste of
any user \(i\) using the preference matrix \(P\) learned by LASSO,
because the columns of \(P_i\) are labeled with tags.

LASSO has been appreciated for its explainability
\autocite{tibshirani1996regression}: the row preferences of each user
are sparse, allowing to capture the tags that explain best the ratings
of every user.

As an example, for a certain user among the authors, LASSO or BALSE
report that his six most preferred tags are: \emph{kneehighs},
\emph{cat}, \emph{serafuku}\footnote{\emph{Serafuku} means ``Japanese
  school uniform''.}, \emph{twin braids}, \emph{japanese clothes} and
\emph{angry} whereas his six most disliked tags are: \emph{pleated
skirt}, \emph{standing}, \emph{silver hair}, \emph{window}, \emph{torn
clothes} and \emph{skirt}. Using this information, LASSO or BALSE can
explain a recommendation: ``We recommend to you the anime \emph{Chivalry
of a Failed Knight}, because there is a girl with \emph{twin braids},
\emph{serafuku} and \emph{japanese clothes}'' or a warning: ``You might
not like the anime \emph{The Asterisk War: The Academy City on the
Water} because there is a girl with a \emph{pleated skirt}, even though
there are \emph{kneehighs} and \emph{serafuku}.''

\section{Conclusion and Future Work}\label{conclusion-and-future-work}

We proposed BALSE, a model for recommending anime and manga that makes
use of information that is automatically extracted from posters. We
showed that our model performs better than the baseline models,
especially in the item cold-start scenario.

This paper is a proof a concept and the obtained results are very
encouraging. Indeed, the blending Steins gate is such that any
improvement made on any block would improve the overall performance of
the approach. As future work, we plan to replace blocks in our
architecture with more complex models: Illustration2Vec could be
replaced with residual networks \autocite{he2016deep}, ALS could be
replaced with factorization machines \autocite{rendle2010factorization}
or co-factorization \autocite{fang2011matrix}, LASSO could be replaced
with Localized Lasso \autocite{yamada2016localized}, a variant that
works well for few samples, many features. We also to integrate more
side information, for instance the drawing style of the image, or tags
coming from open databases such as AniDB\footnote{http://anidb.net} or
AniList\footnote{https://anilist.co}, in order to improve the
explanation of the users' preferences.

For the sake of simplicity, we mapped the categorical ratings like,
dislike, etc. to ad-hoc values, but we could instead use ordinal
regression methods \autocite{pedregosa2017consistency}. However, they
require more computation to be trained properly. Ensemble methods that
blend more than two models could be considered
\autocite{sill2009feature}, or that rely not also on the number of
ratings provided for a certain item, but on the number of ratings
provided by a certain user, or the number of works that contain a
certain tag.

Here, we mitigated the problem of item cold-start recommendation through
the use of extra information on the item side. Obviously, similar
results could be obtained for the user cold-start problem, provided
enough data is available to describe the users.

Using BALSE, recommender systems can automatically replenish their
database, where new items go through the tag prediction track and the
explainable model in order to justify the recommendations for their
first users, and automatically go to the main track when sufficient
ratings have been collected.

\section*{Acknowledgments}\label{acknowledgments}
\addcontentsline{toc}{section}{Acknowledgments}

This work was carried out while Florian Yger was a visiting researcher
at RIKEN Center for AIP, and \KevinCocchi{} and Thomas Chalumeau were
interns at Mangaki. We would like to thank Nicolas Hug and
\EtienneSimon{} for their helpful comments and \Sedeto{} for kindly
providing the example illustration of Figure \ref{illustration2vec}.

\printbibliography


\end{document}